\documentclass[journal,comsoc]{IEEEtran}
\usepackage{amssymb}
\usepackage{hyperref}

\usepackage{url}
\usepackage{amsthm}
\usepackage{multirow}
\usepackage{bbm}
\usepackage{caption}
\usepackage{mathtools}
\usepackage{array}
\usepackage{algorithm}
\usepackage{algpseudocode}
\usepackage{amsmath}
\usepackage{verbatim}

\usepackage{graphicx} 
\usepackage{multicol}
\usepackage{epstopdf}
\usepackage{cite}
\usepackage{float}
\usepackage{enumitem}
\usepackage{array}
\usepackage{lipsum}
\usepackage{fixltx2e}
\usepackage{amsfonts}
\usepackage{algpseudocode}%
\usepackage[numbers,square, comma, sort & compress]{natbib}
\ifCLASSOPTIONcompsoc
\usepackage[caption=false,font=normalsize,labelfont=sf,textfont=sf]{subfig}
\else
\usepackage[caption=false,font=footnotesize]{subfig}
\fi

\ifCLASSINFOpdf
 \else
  \fi
\usepackage{amsmath}

\usepackage[cmintegrals]{newtxmath}
 \usepackage{multirow}
 \usepackage{graphicx}
 \usepackage{lscape}
 \newcommand{\de}{$\rm{dE}^3$}
\newcommand{\dets}{$\rm{dE}^3$-$\rm{TS}$}

\hypersetup{
    colorlinks=true,
    urlcolor = blue
    }
    
\begin{document}

\title{Distributed Learning in Ad-Hoc Networks: A Multi-player Multi-armed Bandit Framework}

\author{Sumit J.~Darak and Manjesh K.~Hanawal
	\IEEEcompsocitemizethanks{\IEEEcompsocthanksitem Sumit J. Darak is with the Department of ECE, IIIT-Delhi, India. E-mail: sumit@iiitd.ac.in. 
	Manjesh K. Hanawal is with IEOR, IIT Bombay, India. E-mail: mhanawal@iitb.ac.in.
}

}

\maketitle

\begin{abstract}
	Next generation networks are expected to be ultra dense with very high peak rate but relatively lower expected traffic per user. For such scenario, existing central controller based resource allocation may incur substantial signaling (control communications) leading
to a negative effect on the quality of service (e.g. drop calls), energy and spectrum efficiency. To overcome this problem, cognitive ad-hoc networks (CAHN) that share spectrum with other networks are being envisioned. They allow some users to identify and communicate in `free slots' thereby reducing signaling load and allowing the higher number of users per base stations (dense networks). Such networks open up many interesting challenges such as resource identification, coordination, dynamic and context-aware adaptation for which {\em Machine Learning} and {\em Artificial Intelligence} framework offers novel solutions. In this paper, we discuss state-of-the-art multi-armed multi-player bandit based distributed learning algorithms which allow users to adapt to the environment and coordinate with other players/users. We also discuss various open research problems for feasible realization of CAHN and interesting applications in other domains such as energy harvesting, Internet of Things, and Smart grids.
\end{abstract}

\section{Introduction}
Next generation wireless networks propose to combine the features of both cellular and ad-hoc networks to support a wide range of new services like, high-speed multimedia, mission-critical control operations, telemedicine, and the Internet of Things  (IoT), which come with diverse service requirements \cite{NR_1}. While the cellular setup aims to offer `guaranteed service' through a dedicated network infrastructure, the ad-hoc setup aims to connect more users to the network with little infrastructure and utilize any spare resources in the network, thus shows promise for better utilization of network resources and provide seamless connectivity. Such heterogeneous networks are expected to greatly enhance user experience and have a huge impact on all aspects of human lifestyles, but they open up many interesting challenges such as resource identification, coordination, dynamic and context-aware network adaptation in the complex environment. {\em Machine Learning} and {\em Artificial Intelligence} framework offers novel solutions to make and adapt to the environment and self configure the network parameters to achieve the best possible performance \cite{ad_hoc_2}.

The cognitive ad-hoc network (CAHN) approach being considered in the next generation networks allows utilization of unlicensed or shared spectrum in addition to the licensed spectrum by the cellular networks, making available more bandwidth for the users. Further, CAHN reduces the signaling (control communications) load allowing a higher number of users supported at the base stations. However, users in an CAHN need to operate in a decentralized fashion due to the lack of common control channels or a central coordinator making efficient utilization of network resources challenging. Moreover, the requirements of each user and its local environmental conditions (congestion, fading, etc.) may be different, which further complicates allocation of resources. In such situations, users not only have to learn about their local environment but also that of the others in the network to efficiently use the network resources and maximize the overall network performance (social utility).

We discuss distributed algorithms that aim to optimize network social utility using learning and coordination mechanisms. The learning will help users to get aware of the local environment and coordination mechanisms involve an exchange of information between the users so that they are aware of the overall network environment. Multi-armed bandits (MAB) provide a standard framework for learning in uncertain environments and are applied extensively  in the study of CAHN.  We are interested in {\em bandit algorithms} augmented with the distributed {\em signaling schemes} that aim to learn local network parameters and allow users to exchange information without requiring a central coordinator or a control channel. The main challenge in developing bandit algorithms for ad-hoc networks is that there could be multiple users in the network trying to use the same set of resources. A collision will occur if more than one user select a channel resulting in a loss of transmission for all the colliding users. Further challenges arise as the network may be dynamic where users join and leave the network any time which complicates the coordination process and the environment may not be stationary, i.e., network parameters such as channel statistics may evolve with time.

In the absence of any central coordinator or a control channel, users collide with each other and through collisions will get to know about the presence of other users in the network. Though collisions result in loss in throughput, they are the only means for information exchange among the users in CAHN. Specifically, an user can collide with others  in a certain pattern to signal a bit sequence which encodes desired information. However, any such communication through collisions results in wastage of spectrum resource as well as battery power which leads to loss in throughput and operational time, respectively. Hence any collision based signaling scheme has to be efficient so that the user terminals can have longer life. Thus, the learning algorithms for next generation wireless networks should incorporate efficient signaling schemes for establishing coordination among the users.

	\begin{table*}[!t]
		\centering
		\caption{Comparison Between Various Radio Models}
		\resizebox{\textwidth}{!}{%
			\begin{tabular}{|p{2.5cm}|l|l|l|p{2.5cm}|p{2.5cm}|p{1.75cm}|p{2.5cm}|p{1.35cm}|}
				\hline
				\textbf{Radio}                                  & 
				\textbf{Type} & \textbf{Sensing} & \textbf{Transmission} & 
				\textbf{Simultaneous Sensing and Transmission} & 
				\textbf{Analog and Digital Front-end Complexity} & 
				\textbf{Baseband Processing Complexity} & 
				\textbf{Orthogonalization, User and channel Estimation} & 
				\textbf{No. of Collisions}\\ \cline{1-9} 
				
\multirow{2}{*}{\textbf{Wideband  (WB)}}&\textbf{I}&WB& Multi-band& Yes & 
Extremely High & Very High& Easy & Few  \\ 
\cline{2-9} 
 & \textbf{II} & WB & Multi-band & No & Extremely High & Very High & Easy  & Few 
 \\ 
 \cline{2-9} 
  & \textbf{III} & NA &Multi-band& Only Transmission & Low & Very Low & 
  Challenging  &Very High \\ 
  \cline{1-9} 
\multirow{2}{*}{\textbf{Hybrid}} & \textbf{I} & WB &Single-band & Yes & Very 
High& Very High & Easy & Few \\
\cline{2-9} 
 & \textbf{II} & WB  &Single-band  & No &Very High &Very High & Easy &  Few \\ 
 \cline{1-9} 
 \multirow{3}{*}{\textbf{Narrowband (NB)}} &\textbf{I}& NB & Single-band  & Yes 
 & Low  &	Low & Very Challenging & High \\
  \cline{2-9} & \textbf{II} & NB & Single-band & No& Low  & Low  & Very 
  Challenging & High \\ 	 
  \cline{2-9}                                           & 
 \textbf{III} & NA  & Single-band & Only Transmission & Extremely Low & 
 Extremely Low & 
 Extremely Challenging &  Very High\\ \hline
			\end{tabular}%
		}
		\label{radio_model}
	\end{table*}
	
The organization of the paper is as follows: In Sections \ref{spectrum} and \ref{radio} we discuss the spectrum and radio models used in the next generation wireless networks. In Section \ref{channel} we discuss the various channel model studied in the literature. We give preliminaries of MAB setup in Section \ref{model} and discuss various existing and proposed algorithms in Sections \ref{SN} and \ref{ASN}. In Section \ref{ORP}, we discuss open research problems followed by conclusions in Section~\ref{conclusion}.


\section{Spectrum Model}\label{spectrum}

Cognitive ad-hoc networks can be deployed in licensed, shared and unlicensed spectrum. 
In February 2018, 3GPP envisioned new radio (NR) based on a revolutionary 
path of spectrum sharing thereby enabling the wide range of deployments 
and services from enhanced mobile broadband to mission-critical services 
to a massive Internet of Things (IoTs). The NR is expected to operate not 
only in licensed spectrum but also in the shared (2.3 GHz/ 3.5 GHz) as 
well as unlicensed spectrum (2.4 GHz / 5-7 GHz / 57-71 GHz). 
{ Other than 
cellular applications, CAHN are also being explored for air-to-ground communications in $L$-band, dedicated short-range communication in 5 GHz 
band and device-to-device (D2D) communication in cellular frequency bands.}

In the licensed spectrum, cognitive users (or secondary users, i.e., SUs) need to avoid interference with the licensed users (or primary users, i.e., PUs) and hence, they must sense the channel before each transmission. In the shared and unlicensed spectrum, users can transmit directly without sensing as there are no PUs. From the 
architecture perspective, radio terminals may not need spectrum sensing unit while operating in unlicensed and shared spectrum which makes them area and power efficient.  However, various works have shown that sensing in the unlicensed spectrum significantly reduces the number of collisions which in turn improves the utilization of the spectrum and hence the throughput \cite{ICL2018_scf,Arxiv2018_Trekking}. From the learning perspective, the design of distributed algorithm and its analysis largely depends on the type of spectrum. In the case of unlicensed spectrum, each user has to learn the quality of channels. Whereas in the licensed spectrum, in addition to the quality of the channels, the users need to learn the probability of the channels being occupied by PUs (occupancy rate) and identify transmission parameters of PUs to choose the channel as well as transmission bandwidth, power, modulation scheme etc. Thus, learning the optimal allocation can be slower in the licensed spectrum.

\section{Radio Model}\label{radio}

In CAHN, each user corresponds to a radio terminal that decides on its own which channel to select for transmission in each time slot. For such decision-making, the amount of information learned and signaled to others depends on the  sensing and transmission capabilities of radio and it impacts the performance and complexity of the distributed algorithm that each user runs on its terminal. Based on these capabilities, the radio terminal can be broadly classified in as Type I, II and III as shown in Table~\ref{radio_model}. Type I radios are more sophisticated and can sense and transmit 
simultaneously. Type III radios are least sophisticated and can only 
transmit with no sensing capability. Type II radios have intermediate capability and can either sense or transmit in a given slot.
Type I radio consists of two independent analog front-ends (AFE), one 
for sensing and other for transmission, each consisting of an antenna, matching 
units, amplifiers, analog-to-digital or digital-to-analog converters etc. In 
Type II and III radios, only one AFE is present. 

The radios of each type can be further classified based on width of the spectrum they can sense. 
The radios with wideband sensing capabilities can sense multiple channels 
simultaneously. The AFE of the wideband sensing is significantly 
complex due to the need for high-speed analog-to-digital converters and 
stringent analog signal processing to satisfy Nyquist sampling criteria. Though there is some progress on sub-Nyquist 
sampling based wideband sensing which exploits the sparsity of the wideband spectrum, it is still computationally intensive and needs 
complex digital baseband processing algorithms. Hence, Type I and II wideband radios are power hungry and may not be suitable for battery operated radio terminals. However, design of distributed learning algorithms with such wideband radios is easy as it allows to gather and signal more information in each time slot which is beneficial to estimate the number of active users and channel quality faster. 

The narrowband radios can operate on a single channel in each time slot. Type I narrowband radio has two narrowband AFE so that simultaneous 
sensing and transmission over two different channels is possible. The Type II narrowband radio allows either sensing or transmission only over single narrowband channel which makes them area and power efficient than wideband radios. They are widely used in the CAHN. The narrowband Type III radio have the least complexity but  there is no way to confirm whether fading or collision lead to the transmission failure. Hence the design of learning and coordination algorithm for Type III narrowband radios is hard and is one of the challenging open research problem.

Radio consisting of both narrowband and wideband AFEs are also available and are referred to as Hybrid radios \cite{ICML16_MultiplayerBandits_RosenkiShamir}. In such radios, the narrowband AFE is used for transmission and the wideband AFE is used for sensing. Type I hybrid radio can simultaneously transmit on a channel and sense all channels. Type II hybrid radio can transmit on a channel or sense all channels in each time slot. 

In Type II radio terminals, the learning algorithms have to switch between the sensing and transmission modes effectively. Longer sensing times reduce the transmission time resulting in reduced throughput. On the other hand, longer transmission times reduce the sensing time causing higher sensing errors that lead to inaccurate learning, false alarms and interference to licensed  users.


	\renewcommand{\arraystretch}{1.5}

\section{Channel Model}\label{channel}
Secondary users can transmit on any channel in unlicensed spectrum, whereas they can do so only on the channels not occupied (vacant) by PUs in the licensed spectrum. Each channel gives a rate/reward when a collision-free transmission occurs on it. The reward obtained on a channel depends on various factors like fading, interference, shadowing etc. The reward process can be stochastic or non-stochastic/adversarial.

In the stochastic setting, the rewards obtained from each channel are assumed to be random and drawn from an arbitrary distribution with bounded support. The reward distribution of each channel is either fixed throughout the horizon (i.e., stationary) or changes at regular interval (i.e., quasi-stationary). The samples obtained from a channel can be independent or Markovian. Most of the works in the literature consider that the samples are i.i.d for analytic tractability while only few consider the Markovian property. When collision occurs on a channel, all the users involved in the collision do not get to observe the reward sample. However, collision allows users to know about presence of other users in the networks and could be deliberately enforced by an algorithm for exchange of information between the users. The users involved in collision may still get some rate/reward when techniques like successive interference cancellation are used. However, most of existing works take the pessimistic approach and any such reward is ignored in the learning process. 

In the adversarial setting, the rewards could be generated in an arbitrary fashion (i.e., non-stationary), and in the worst case, could be generated by an adversary that aims to minimize the reward obtained by the users.

\section{Multi-Armed Bandit Model}\label{model}
In this section we discuss the Multi-player multi-armed bandits (MPMAB) setting that is widely used to study learning in CAHN. MPMAB is a variant of the standard MAB \cite{Book12_RegretAnalysis_Bubeck} where multiple players aim to maximize sum of their rewards (social reward) over the same set of arms. In this setting, the players do not communicate with each other and may not know number of other players in the game. If two or more players select the same arm simultaneously, they experience `collision'  and none of them receive any reward. Further, a player experiencing a collision will not know with whom and how many she collided. The performance of an algorithm in MPMAB is measured in terms of the amount of regret it incurs, where regret is defined as the difference between 1) highest total network reward (summed across all users) obtained by an oracle that knows all the parameters (mean rewards, number of users) using a central controller and 2) total reward obtained by the distributed algorithm without the knowledge network parameters and without a central controller. An oracle can achieve highest total reward by assigning a distinct arm to each player in the top $N$ arms, i.e., optimal allocation. Top $N$ arms correspond to the first $N$ arms when they are listed in decreasing order of their mean values. Any algorithm that achieves sub-linear regret gives the same average reward per round asymptotically as that achievable by an optimal allocation. A linear regret implies that the algorithm has not been able to learn the optimal allocation. The aim in MPMAB is to design distributed algorithms that have small sub-linear regret.  


The setting of MPMAB well models the CAHN where multiple users transmit on a common set of channels without any communication among them and without any central coordinator. The users and channels of the networks correspond to the players and arms in MPMAB, respectively. The reward from a channel is the rate obtained on it when a collision free transmission happens. 
The MPMAB thus provides the required `distributed learning and coordination' framework for CAHN. The optimal social reward in the case of unlicensed spectrum depends on the number of players and the mean reward observed by each player on each channel. For the licensed case, in addition to these parameters, channel occupancy rates also affect the optimal network reward.
\begin{figure*}[!h]
\centering
			\includegraphics[scale=0.65]{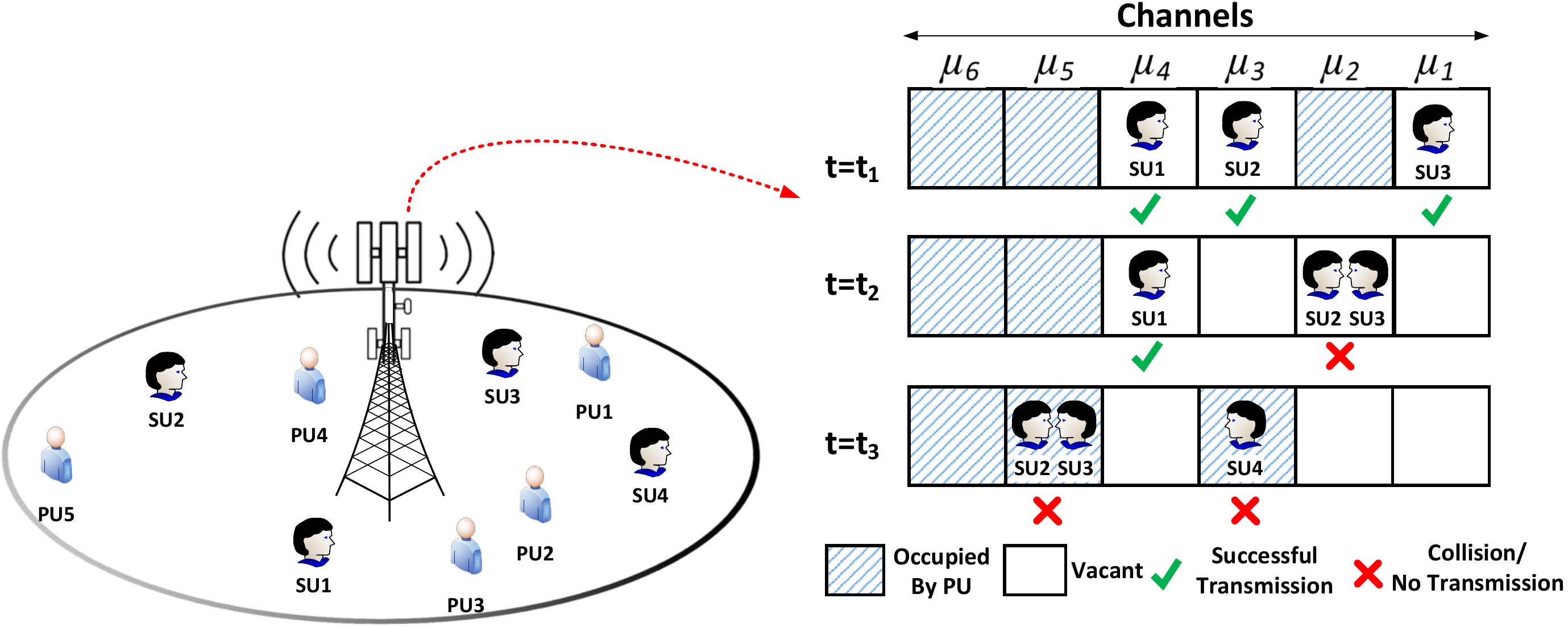}
		 	\caption{CAHN consisting of PUs and SUs. Active users transmit data to intended receiver via base station (In D2D or M2M communications, users communicate directly with each other). Base station allocates channel only to PUs while SUs need to select their channels independently. Corresponding channel selection and collision scenarios are indicated. In unlicensed spectrum, PUs are not present.}
		 	\label{network}
		 \end{figure*}
		 

A typical scenario of a network consisting of $5$ PUs denoted as $PU_i, i=1,2,3,4,5$, and $4$ SUs denoted as $SU_i, i=1,2,3,4$, is shown in Fig.~\ref{network}. The network consists of $6$ channels and mean rate observed by each user on channel $i=1,2,\cdots,6$ is denoted as $\mu_i$. At time $t=t_1$,  channel $6,5,2$ are occupied by PUs and each of the other vacant channels ($4,3,2$) are used by the SUs without any overlap and hence all their transmissions are successful. At $t=t_2$, both $SU_2,SU_3$ select the vacant channel $2$ and incur collision, whereas transmission of $SU_1$ is successful. Time $t=t_3$ depicts the scenario where all the SUs select occupied channels and do not transmit.

Several authors have considered the problem of distributed learning in CAHN under varied network scenarios. They can be broadly classified as follows:
\begin{itemize}
	\item The channel statistics are the same for all users (homogeneous) or  they are different (heterogeneous)
	\item Number of users in the network remains fixed (static) or can change (dynamic)  
\end{itemize}
Further, in each of these cases number of users in the network can be known or unknown. The scenario becomes more challenging as we move from a static-homogeneous network with known number of users to a dynamic-heterogeneous network with unknown number of users. In the following sections, We discuss each of these scenarios and state-of-the-art algorithms. Sensing capability of the users also play role in the learning process-- it becomes simpler with wideband sensing radios compared to that with narrowband sensing radios, however the cost is higher in the former case. Most of the works in the literature consider the narrowband sensing and is assumed to the default option in the subsequent discussion unless stated otherwise.  




\section{Homogeneous Networks: Static and Dynamic}\label{SN}

 In homogeneous-static networks, mean rate experienced on a channel is same for each user. 
 Let $N$ denote the number of users in the network. $N$ remains fixed in the static networks and it can keep changing in the dynamic network. Various algorithms have been proposed to learn channel statistics, estimate the number of active users ($N$) and orthogonalize them on the optimum channels top $N$ channels.

The $\rho^{\mbox{{\tiny RAND}}}$ \cite{JSAC11_DistributedLearning_Anadakumar} algorithm is one of the first distributed algorithm which considered the homogeneous-static setting with a known $N$. $\rho^{\mbox{{\tiny RAND}}}$ algorithm uses well known upper confidence bound (UCB) based MAB algorithm to learn channel statistics while using random reordering mechanism for orthogonalization. Though $\rho^{\mbox{{\tiny RAND}}}$ offers asymptotic logarithmic regret, it incurs large number of collisions due to random reordering where users randomly select new channel after every collision which in turn may results in collision with other users. The musical chair based MCTopM algorithm in \cite{ALT2018_MultiplayerBandits_BessonKaufma} overcomes this drawback and it is the current state-of-the-art algorithm for homogenoeus-static networks with known $N$. In MCTopM, when two users collide, users switch to new channel only when current channel is non-optimal and user is not locked. This allows faster orthogonalization after collision. For the case of an unknown $N$, modified $\rho^{\mbox{{\tiny EST}}}$ is also given in \cite{JSAC11_DistributedLearning_Anadakumar}, but its guarantees hold only asymptotically.  All these algorithms work only for the static case and cannot extend to the dynamic scenarios.
Other variants of stochastic MPMAB also consider a fixed $N$ and address the issue of fairness \cite{TSP14_DistributedStochastic_GaiKrishnamachari}, \cite{JSAC2014_DesignAnalysis_ZhangYao}. In \cite{TSP14_DistributedStochastic_GaiKrishnamachari}, authors aim to achieve fairness in throughput obtained for the users using modified UCB based algorithm whereas fairness is achieved in \cite{JSAC2014_DesignAnalysis_ZhangYao} by allowing each user to access each channel for the same proportion of time via sequential hopping (SH) based approach.



The works in \cite{KDD14_ConcurrentBandits_AvnerMannor} and \cite{ICML16_MultiplayerBandits_RosenkiShamir} consider homogeneous-dynamic network with unknown $N$. The MEGA algorithm in \cite{KDD14_ConcurrentBandits_AvnerMannor} uses the classical $\epsilon$-greedy MAB algorithm and ALOHA based collision avoidance mechanism. Though collision frequency reduces in MEGA as time increases, it may not go to zero as shown in \cite{ICML16_MultiplayerBandits_RosenkiShamir}. To overcome this, \cite{ICML16_MultiplayerBandits_RosenkiShamir} develops MC algorithm that incurs collisions due to random hopping (RH) in the initial learning phase and guarantees collision-free access over optimum channels subsequently. Though MC performs better than MEGA, its performance in the learning phase is poor -- MC uses collision information to estimate $N$ and forces a large number of collisions to get a good estimate. In \cite{ICL2018_scf}, we proposed secondary user co-ordination with fairness (SCF) algorithm for static network with unknown $N$ and extended it to dynamic networks (dynamic SCF i.e., DSCF) using epoch approach.  
The proposed algorithms are based on novel channel hopping
and one-hot sensing (OHS) approaches. In hopping approach,
the users are orthogonalized via RH and then
follow collision-free SH which helps to estimate channel statistics. In the OHS approach, each user estimates $N$ by sensing the number of SH users. The SCF algorithm is the current state-of-the-art algorithm for static network.
For dynamic networks, though DSCF algorithm outperforms MEGA and dynamic MC (DMC) algorithms, use of epoch approach makes it far from being optimal due to repetitive RH and SH phases.

Our Trekking for Static Network (TSN) for static and its variant Trekking for Dynamic Network (TDN) in \cite{Arxiv2018_Trekking} overcomes the above drawbacks.
As opposed to existing algorithms which separate estimation and orthogonalization tasks, we show that users can be settled on the top $N$ channel without knowing $N$. Specifically, TSN and TDN are based on novel trekking approach where users operating on a channel always looks to operate on a vacant channel with better quality.
Thus, all the users end up transmitting on the optimum channels
without knowing $N$. For dynamic case, we do not need epoch approach and analysis shows that it outperforms DMC and DSCF algorithms.

For illustration, we compare the performance of these algorithms for static and dynamic networks. For static network, we consider MCTopM in \cite{ALT2018_MultiplayerBandits_BessonKaufma} and its variant with unknown $N$ named as UMCTopM (assumes $N=K$), SH in \cite{JSAC2014_DesignAnalysis_ZhangYao}, TSN and SCF algorithms. We set number of channels $K= 8$, $N\in \{4,8\}$ and mean rates of channels as 
$\mu=\{0.29,0.36,0.43,0.50,0.57,0.64,0.71,0.78\}$. All the numerical results have been shown after averaging the values obtained over $50$ independent experiments. Various parameters of algorithms are chosen carefully to achieve best possible performance in terms of average regret. 

\begin{figure*}[!h]
\centering
			\includegraphics[scale=0.6]{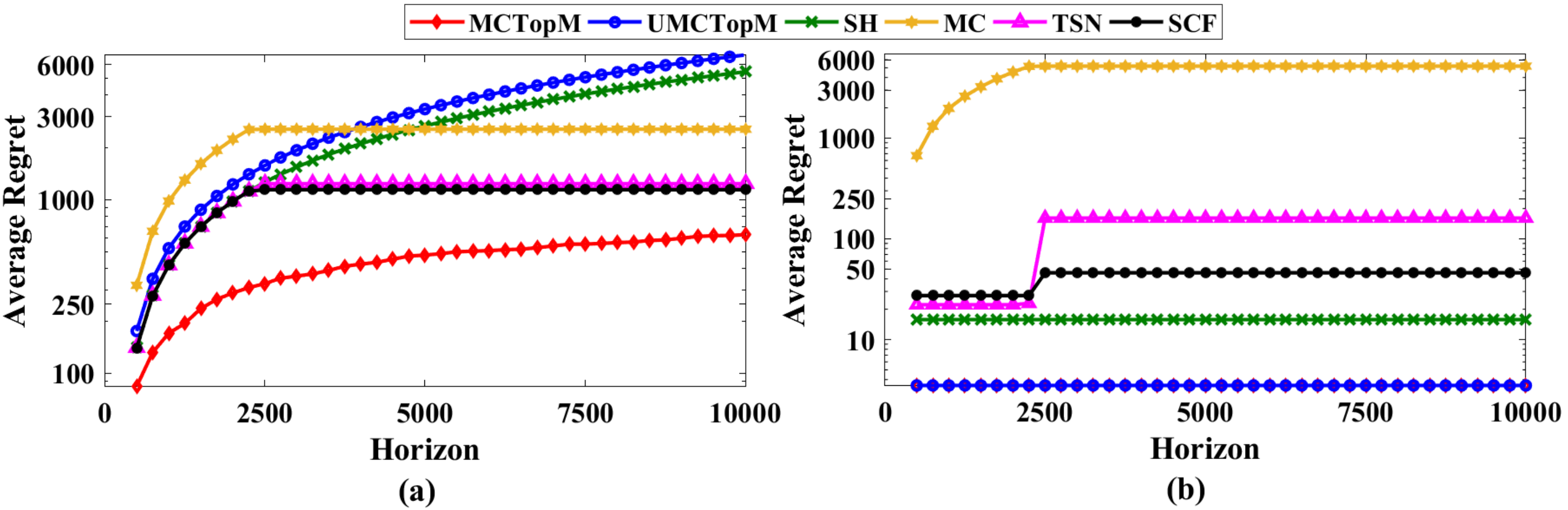}
		 	\caption{Average regret at different instants of the horizon for static homogeneous network. We assume $K = 8$ with (a) $N=4$, and (b) $N=8$. $y-axis$ is shown on the logarithmic scale.}
		 	\label{static_S_u48}
		 \end{figure*}

Fig.~\ref{static_S_u48}(a) shows the experimental results for the static network with $N=4$. It is evident that UMCTopM and SH algorithms perform poorly as due to frequent selection of non-optimal channels. Also, we can see that the TSN and SCF offer lower regret than MC. The improvements is because of the collision-free SH in comparison to RH based learning in MC algorithm. As expected, MCTopM has lower regret than the SCF due to the knowledge of $N$ while the others do not have this knowledge.  The regret plot with constant slope in TSN, SCF and MC algorithms indicate the orthogonalization of the users in the top $N$ channels leading to zero regret after some initial rounds.

\begin{figure}[!h]
\centering
			\includegraphics[scale=0.5]{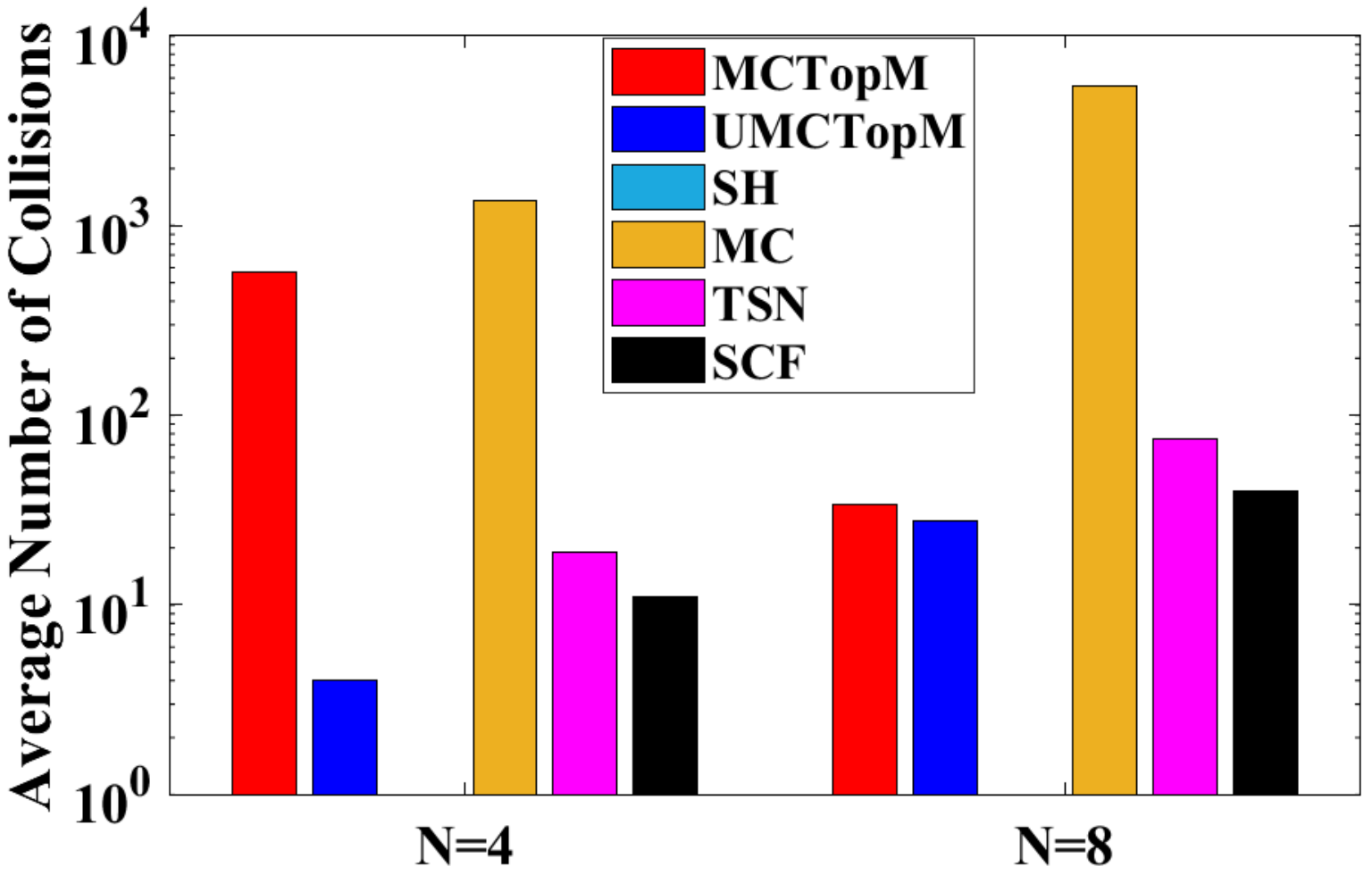}%
		 	\caption{Average number of collisions for static homogeneous network with $K = 8$ and $N=\{4,8\}$. $y-axis$ is shown on the logarithmic scale.}
		 	\label{static_S_c}
		 \end{figure}

Fig. \ref{static_S_u48}(b) shows regret plots for the case of saturated network, i.e.,  $K=N=8$. As expected, regret of MCTopM and UMCTopM are identical. The performance of the SH algorithm improves with $N$ due to
selection of all the channels uniformly via collision-free SH approach. Note that the regret of TSN and SCF algorithms improves as $N$ increases while that of MC algorithm degrades significantly due to a large number of collisions. From collision perspective, as shown in Fig. \ref{static_S_c}, SCF and TSN algorithms offer fewer collisions than others except SH and UMCTopM algorithms. However, regret performance of SH and UMCTopM are significantly poor.

\begin{figure}[!h]
\centering
			\includegraphics[scale=0.5]{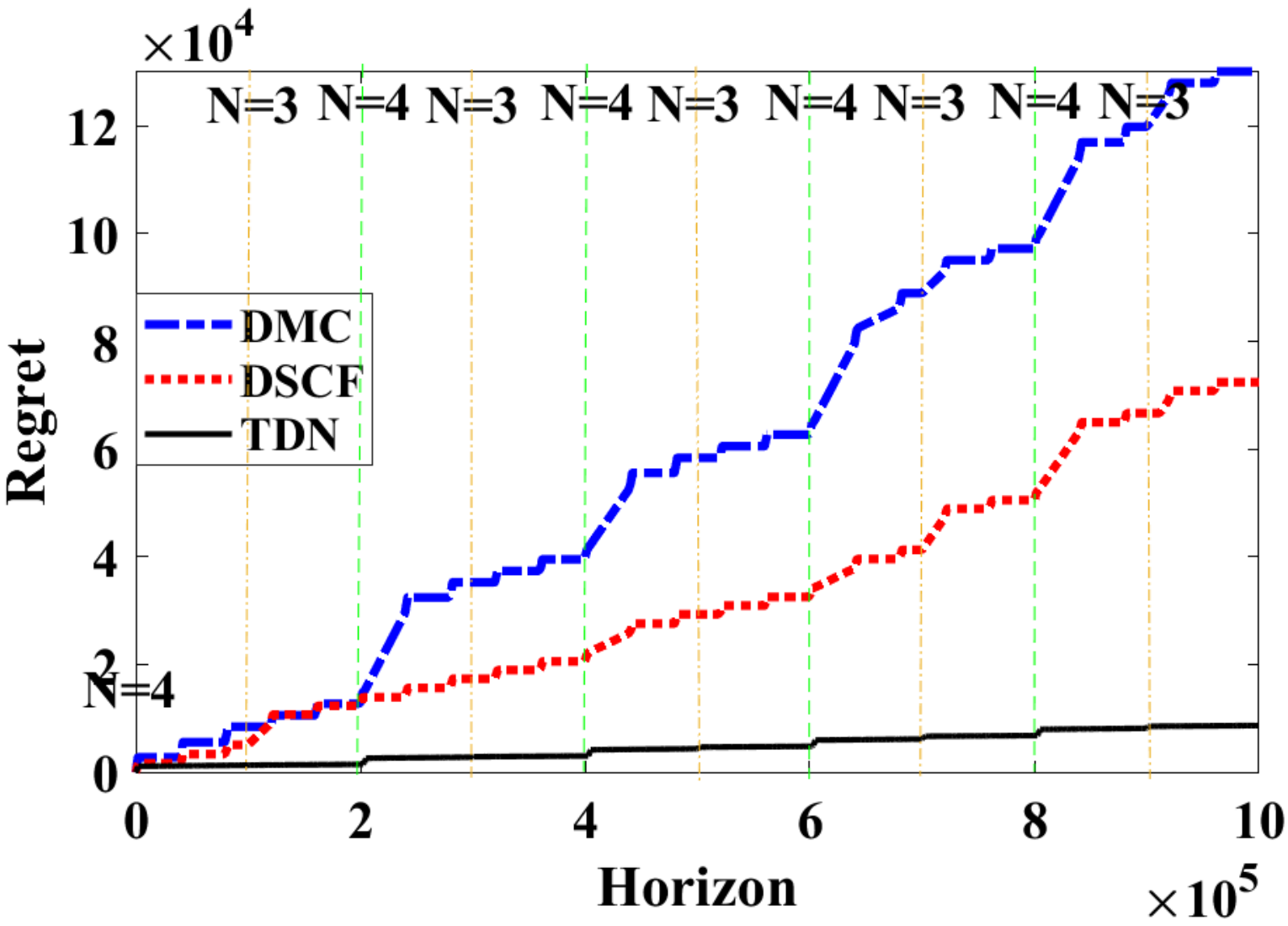}%
		 	\caption{Average regret at different instants of the horizon for dynamic homogeneous network for $K = 8$ and stationary channel statistics.}
		 	\label{dyn_s}
		 \end{figure}

		 Next, we compare the performance of the DMC, DSCF and TDN algorithms for dynamic networks in Fig.~\ref{dyn_s}. We mark the time of entry and exit of the user with a green dashed and yellow dot-dashed lines, respectively. We start with four users and at every 100000-time slots, we alternate between the user leaving and entering the network. As expected, DSCF algorithm performs better than the DMC and the difference increases with time. Though epoch based algorithms allow users to adapt the channel statistics over time, they incurs regret in each epoch irrespective of whether the user leaves or enters the network. The epoch-free TDN algorithm avoids such regret and hence significantly outperforms others. Additional simulation results considering various scenarios for static and dynamic networks are given in \cite{Arxiv2018_Trekking}.
		 
\section{Heterogeneous: Static and Dynamic}\label{ASN}
In heterogeneous networks, in addition to unknown statistics, channels are
asymmetric across the users, i.e., average throughput offered
by channels may not be the same for all users due to their geographical separations. In such scenarios, to achieve optimal network throughput users not only need to learn the channel statistics experienced by them but also that experienced by the others. This is challenging problem in CAHN and only few works works have addressed it. 


 The Coordinated
 Stable Marriage Multi-Armed Bandit (CSM-MAB) algorithm in \cite{INFOCOM16_MultiUserLax_AvnerMannor} focuses on stable orthogonal configuration (SOC) without the need of direct communication between users. In SOC, algorithm is said to have converged if no user will have an incentive to request any other user for a swap of their channels, hence channel switches/swaps will not occur. But SOC may not always lead to optimal configuration. Also, CSM-MAB assumes users are quipped with wideband radios that allows them to know the which channels are selected by other users in each time slot. As wideband sensing more power than the narrowband sensing, CSM-AMB is not suitable for battery operated users. 




The authors in \cite{TIT14_DecentralizedLearning_KalthilNayyarJain},\cite{TCNS2018_DecentralizedLearning_KalthilNayyarJain} consider optimal allocation in the heterogeneous-static network. The $\mbox{dUCB}_4$ algorithm in  \cite{TIT14_DecentralizedLearning_KalthilNayyarJain}  uses Bertsekas' auction mechanism where users to negotiate unique channel.  The authors in \cite{TCNS2018_DecentralizedLearning_KalthilNayyarJain} propose exponentially spaced exploration and exploitation algorithms, named \de and \dets \;, to achieve near-optimal logarithm regret and they outperform $\mbox{dUCB}_4$ algorithm. 

In \cite{INFOCOM2019_optimal_matching}, we develop Explore-Signal-Exploit Repeat (ESER) and its variant named modified ESER (mESER). These algorithms run in phases. In each phase, users explore the channels initially via SH approach and exchange their estimates with other users via novel signaling scheme. After this, users
compute the optimal channel allocation by applying Hungarian method  using the estimated mean rewards and lock on their assigned channel for double the time than in the previous
phase. We also show that mESER, which optimizes the duration of signalling phase, achieves logarithmic regret while ESER, \de and \dets\;  \cite{TCNS2018_DecentralizedLearning_KalthilNayyarJain} algorithms achieve near-logarithmic regret. To the best of our knowledge, is the only algorithm that guarantees logarithm regret without requiring
any problem-specific information (sub-optimality gap).

\begin{figure*}[!h]
	\centering
	\subfloat[]{\includegraphics[scale=0.42]{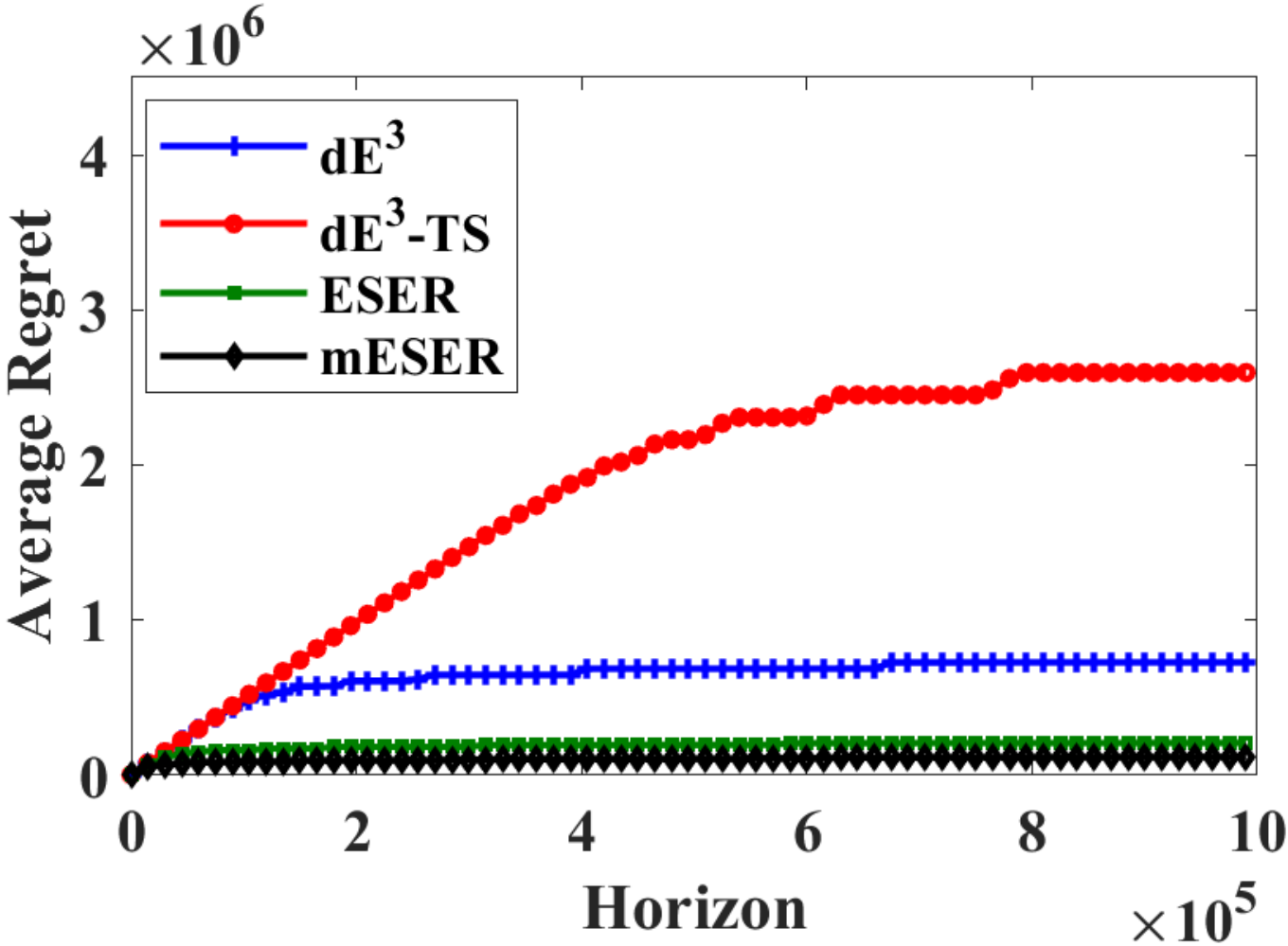}
		\label{fig:}}
	\subfloat[]{\includegraphics[scale=0.42]{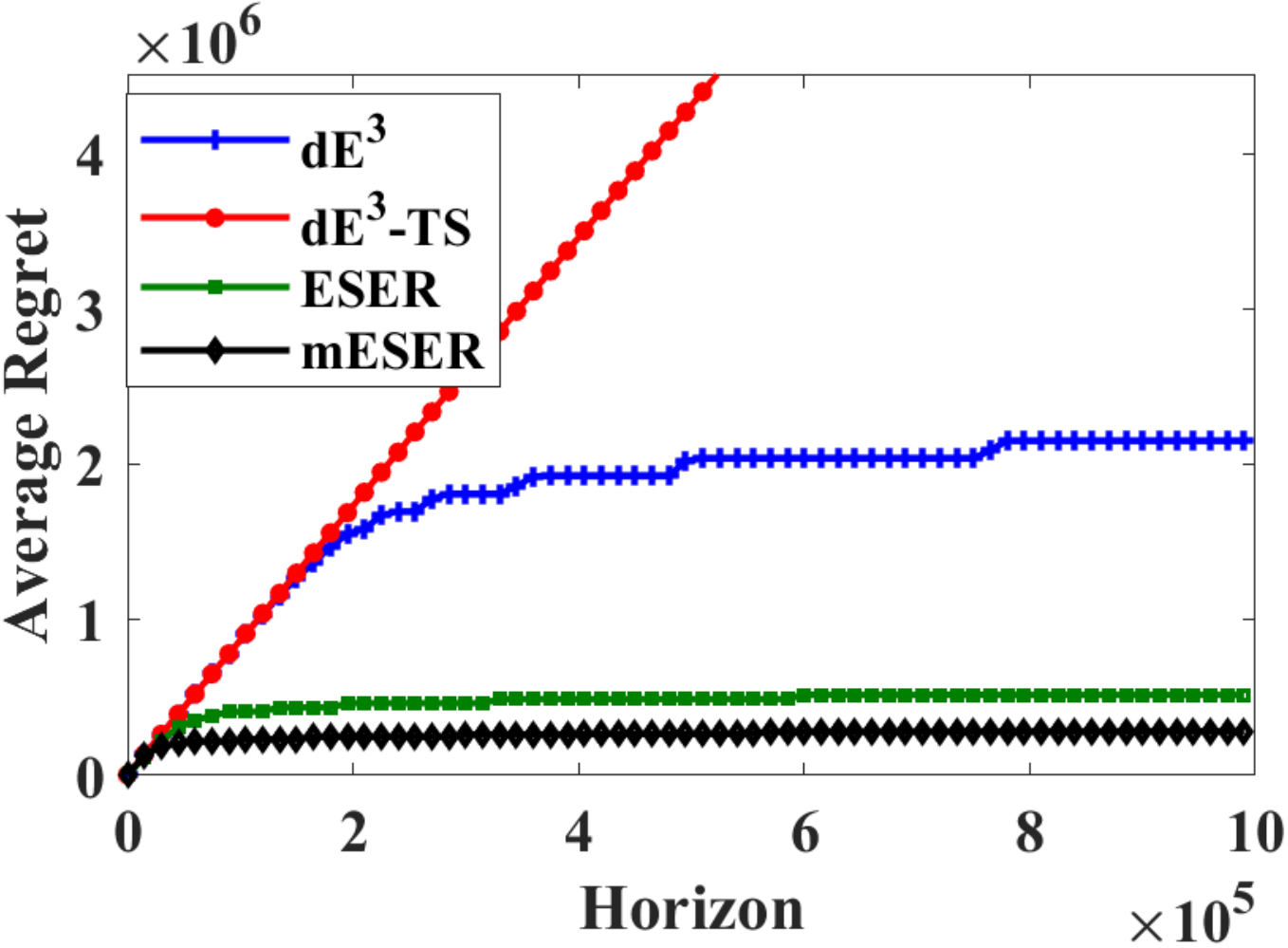}
		\label{fig:}}
	\subfloat[]{\includegraphics[scale=0.42]{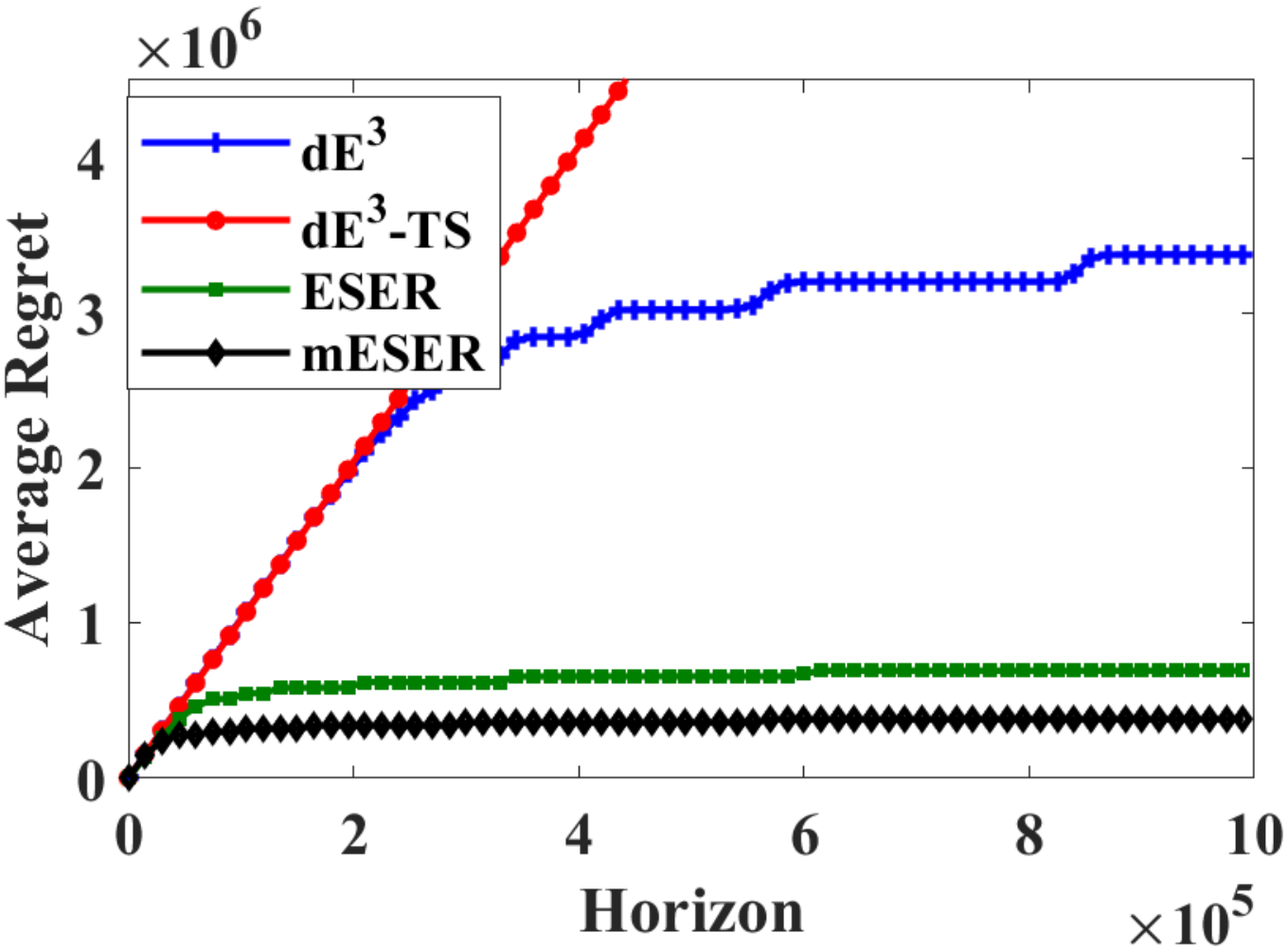}
		\label{fig:}}
	\caption {Average regret comparison for static heterogeneous network. We set $K=12$ and $N=\{6,10,12$\} in (a), (b), (c), respectively.}
	\label{fig:ESERvdE3RegretN}
\end{figure*}

In Fig.~\ref{fig:ESERvdE3RegretN}, we compare cumulative regret of \de, \dets \;, ESER and mESER for $K=12$ and $N=\{6,10,12\}$. As seen, mESER and ESER perform significantly better than both  \de and \dets algorithms.   
When $N$ increases, the performance of our algorithms improve significantly.

The extension of the current algorithms for the heterogeneous-static networks to the dynamic case is not straightforward and it demands more direct and frequent communication between the users. Increased amount of information exchange between the users leads to increase in signaling overheads which may not be desirable in some time critical services and applications. Further, increased communication between the users can be vulnerable to  intentional jamming and Denial- of-Service attacks. 

\section{Open Research Problems}\label{ORP}
Various setups considered in the literature and their limitations are discussed in previous sections. Here, we discuss open research problems and challenges that have not been explored yet for CAHN. 

In existing algorithms give their guaranteed performance assuming that all the users will the run the algorithms faithfully and aim to maximize the network throughput. However such assumption may not always hold due to the decentralized nature of the CAHN. In the worst case, some malicious users might launch Denial-of-Service attack and degrade network performance. It is thus important that the
distributed algorithms are capable of identifying such malicious behavior and guarantee best possible throughput in their presence.

The feedback model in the existing algorithms is real time and independent for each transmission. In resource constrained networks, feedback might get delayed and algorithms need to estimate the average delay. Furthermore, feedback may be aggregated such that users may receive only one feedback for transmissions spanned over multiple channels and time slots. From scheduling perspective, existing algorithms do not consider age-of-information which is a recently proposed measure of freshness or timeliness of the sensed information in applications such as autonomous vehicles, smart grid and transportation systems.

From energy-efficiency perspective, self-sufficient radio terminals are being envisioned which can harvest RF energy and use it later for data transmission. For such radios, algorithms have an additional task of deciding when to transmit and/or harvest and choose the appropriate channel in each case. For multi-user ad-hoc networks, appropriate mode selection for each user is also an interesting research problem. Existing algorithms need sensing hardware to detect the presence of other users and to communicate the information via signalling scheme. Recently, few authors have explored the design of distributed algorithms which do not need sensing hardware making them suitable for battery operated sensor nodes in applications such as IoTs. However, more efforts are needed to improve the performance of such algorithms. 

Rendezvous is the process through which the transmitter and receiver tune themselves to communicate on an identical channel.  Distributed algorithms must take rendezvous cost into account especially for applications such as M2M and D2D communications where users communicate directly with each other. Establishing rendezvous is challenging in the ad-hoc network due to lack of  dedicated control channel between transmitter and  receiver and MAB based channel selection.

From MAB seetings perspective, existing algorithms assume that the environment is stochastic and stationary. However, in real life situations the environment may not be stationary as the background environment can change, especially in the dynamic case. In the case of networks involving primary users, the back ground environment can depend on the primary user traffic pattern and can change arbitrarily. Adversarial MAB setting where no prior assumption is made on the reward distribution is not yet studied for CAHN and is an desirable model to explore.

There has not been significant progress on the mapping of MAB algorithms to architectures for efficient implementation on processor or hardware such as FPGAs or ASICs. This is important since radio terminals must be realized on the hardware unlike applications such as web advertising, medical diagnosis where software realization of the MAB algorithms is sufficient. Among existing MAB algorithms, UCB algorithm and its variants seem hardware friendly as they need simple logarithmic, division and square-root operations. However, implementation of KL-UCB, Thompson Sampling and Bayes-UCB algorithms is challenging due to inherent randomization and optimization approaches for which efficient architectures need to be developed. Another major issue is that existing MPMAB algorithms assume global clock synchronization between users. This significantly increases the radio complexity due to the need of high quality oscillators and stringent synchronization constraints. Thus, algorithms which allow users to automatically synchronize with the network needs to be investigated.


\section{Conclusions} \label{conclusion}
In this paper, multi-player multi-armed bandit (MPMAB) based learning framework and signalling schemes for learning and coordination tasks in cognitive ad-hoc networks are discussed. We considered static  or dynamic and homogeneous or heterogeneous networks where number of users as well as channel statistics are unknown and may vary with time. In each case, existing state-of-the-art algorithms, their advantages and drawbacks are presented along with simulation results. Various open research problems to overcome drawbacks of existing algorithms as well as their hardware implementation are discussed. Applications of MPMAB framework in domains such as RF energy harvesting, smart grid, IoTs etc. makes it an exciting research area enabling smart networks.

\bibliographystyle{IEEEtran}
\bibliography{biblio}


\begin{IEEEbiographynophoto}
{Sumit J. Darak}
received his Bachelor of Engineering (B.E.) degree in Electronics and Telecommunications Engineering from Pune University, India in 2007, and PhD degree from Nanyang Technological University (NTU), Singapore in 2013.
He is currently an Assistant Professor at IIIT-Delhi, India. Dr. Sumit has been awarded India Government’s ‘DST Inspire Faculty Award’ which is a prestigious award for young researchers under 32 years age. He has received \textit{Best Demo Award} at CROWNCOM 2016, \textit{Young Scientist Paper Award} at URSI 2014 and 2017, \textit{Best Student Paper Award} at IEEE DASC 2017 and Second-best poster award in COMSNETs 2019. His current research interests include the reinforcement learning algorithms and reconfigurable architectures for applications such as wireless communications, energy harvesting etc.
\end{IEEEbiographynophoto}

\begin{IEEEbiographynophoto}
{Manjesh K. Hanawal}
received the M.S. degree in ECE from the Indian Institute of Science, Bangalore, India, in 2009,
 and the Ph.D. degree from INRIA, Sophia Antipolis, France, and the University of Avignon, Avignon, France, in 2013. After spending two
 years as a postdoctoral associate at Boston University, he is now an Assistant Professor in Industrial Engineering and Operations Research
 at the Indian Institute of Technology Bombay, Mumbai, India. His research interests include communication networks, machine learning
 and network economics.
\end{IEEEbiographynophoto}

\end{document}